\documentclass[11pt,a4paper]{article}
\usepackage[utf8]{inputenc}
\usepackage[T1]{fontenc}
\usepackage{lmodern}
\usepackage{graphicx}
\usepackage{booktabs}
\usepackage{amsmath,amssymb}
\usepackage{caption}
\usepackage{hyperref}
\usepackage{array}
\usepackage{geometry}
\usepackage{float}

\title{A Hybrid Deep Learning and Anomaly Detection Framework\\
for Real-Time Malicious URL Classification}

\author{
\begin{tabular}{c}
\textbf{Berkani Khaled}$^{1,\star}$ \\
\texttt{bberkani109@gmail.com}
\end{tabular}
\quad\quad
\begin{tabular}{c}
\textbf{Zeraoulia Rafik}$^{2}$ \\
\texttt{zeraoulia@univ-dbkm.dz}
\end{tabular}
}

\date{\today}

\begin{document}
\maketitle

\begin{center}
$^{1}$University of Batna 2, Algeria \\[4pt]
$^{2}$Djilali Bounaama University of Khemis Miliana,\\
Laboratory of Pure and Applied Mathematics (C1151600)\\[6pt]
\end{center}

% Corresponding author note (arXiv style)
\begin{small}
\noindent $^{\star}$Corresponding author: \texttt{bberkani109@gmail.com}
\end{small}

% Abstract follows...

\begin{abstract}
Malicious URLs remain a primary vector for phishing, malware, and cyberthreats. This study proposes a hybrid deep learning framework combining HashingVectorizer n-gram analysis, SMOTE balancing, Isolation Forest anomaly filtering, and a lightweight neural network classifier for real-time URL classification. The multi-stage pipeline processes URLs from open-source repositories with statistical features (length, dot count, entropy) achieving $O(NL+EBdh)$ training complexity and 20ms prediction latency. Empirical evaluation yields 96.4\% accuracy, 95.4\% F1-score, and 97.3\% ROC-AUC, outperforming CNN (94.8\%) and SVM baselines by 50-100$\times$ speedup (Table~\ref{tab:comp-complexity}). A multilingual Tkinter GUI (Arabic/English/French) enables real-time threat assessment with clipboard integration. The framework demonstrates superior scalability and resilience against obfuscated URL patterns.
\end{abstract}


\section{Introduction}
The number and sophistication of cyberthreats have increased along with the internet's exponential expansion, especially those that are spread by bad URLs. A variety of assaults, such as phishing, drive-by downloads, command-and-control communications, and data exfiltration, are launched using malicious websites. Because attackers are constantly changing URLs to avoid detection, traditional blacklisting techniques are unable to keep up with the dynamic and hostile character of contemporary threats. As a result, intelligent algorithms that can recognize intricate patterns in URLs and instantly identify malicious ones have become crucial components of contemporary cybersecurity protection designs \cite{reyes2024,aljabri2022}.

Because machine learning (ML) and deep learning (DL) approaches can identify non-linear relationships in input data and generalize from observed patterns, they have shown considerable promise in the field of malicious URL detection \cite{ha2023,raschka2019}. But there are still a number of obstacles to overcome: class imbalance (lack of labeled malicious data compared to benign URLs); attackers' adversarial techniques that produce highly obfuscated or anomalous URLs that undermine the effectiveness of traditional classifiers; and the majority of detection systems are restricted to monolingual user interfaces and lack real-time usability features.

In order to overcome these problems, this work suggests a hybrid threat detection framework that integrates anomaly filtering via Isolation Forest, deep neural network classification, and character-level feature extraction, all of which are housed within an intuitive, multilingual graphical user interface (GUI). Our method starts by gathering harmful URLs from threat intelligence feeds that are accessible to the public and creating fake benign URLs using popular domain patterns. Then, in order to efficiently handle high-dimensional sparse data, rich characteristics are generated from URLs using both statistical analysis (e.g., length, dot count, entropy) and vectorized token representations via HashingVectorizer.

We use SMOTE (Synthetic Minority Over-sampling Technique), which creates new minority class samples based on preexisting ones, to lessen class imbalance. Furthermore, before training the model, we use Isolation Forest to identify and eliminate noisy or unusual samples, which greatly enhances generalization performance. Using the Keras API from TensorFlow, a fully connected neural network learns from the processed data to categorize URLs as benign or harmful \cite{cerulli2023}.

\section{Related Works}
Researchers in the fields of data mining, machine learning, and cybersecurity have been actively studying malicious URL detection. Static techniques like blacklists and heuristic rule-based systems were the mainstay of early detection tools. Although blacklist-based methods (like PhishTank and Google's Safe Browsing API) are effective against known threats, they have limited generalization and are unable to identify zero-day or quickly changing URLs. Despite being interpretable, rule-based approaches are vulnerable to evasion strategies like polymorphism and URL obfuscation \cite{patil2018}.

Recent studies have used machine learning (ML) models to get over these restrictions by approaching the detection of malicious URLs as a binary classification task. Popular methods that use manually engineered features like lexical patterns (e.g. length, presence of IP addresses, special characters), host-based features (e.g. WHOIS data), and content-based attributes (e.g. HTML structure) include Support Vector Machines (SVM), Random Forests, and Naïve Bayes. Although these approaches perform noticeably better than static strategies, they are constrained by their dependence on feature quality and susceptibility to concept drift in hostile settings \cite{abu2007,cortes1995}.

More sophisticated methods automatically extract features at the character or word level from raw URLs using deep learning (DL), specifically Recurrent Neural Networks (RNNs) and Convolutional Neural Networks (CNNs). These models have proven to be quite accurate and do away with the necessity for manual feature engineering. However, they are less feasible for lightweight or real-time deployments since they frequently call for expensive computational resources and substantial amounts of tagged data \cite{waheed2022,mercaldo2022}.

\section{Proposed System}
The proposed system presents an integrated and robust architecture for malicious URL detection that combines advanced machine learning techniques with a user-centric graphical user interface. The framework is designed to operate efficiently in real-time environments while maintaining high detection accuracy across diverse and imbalanced datasets. Our system addresses the key challenges in URL classification, including feature sparsity, data imbalance, concept drift, and practical deployment.

\subsection{System Overview}
The system is composed of the following key components:
\begin{itemize}
\item Data Acquisition and Preprocessing
\item Feature Extraction with HashingVectorizer
\item Class Balancing using SMOTE
\item Outlier Detection via Isolation Forest
\item Deep Neural Network Classifier
\item Multilingual Graphical User Interface (GUI)
\end{itemize}

Each of these modules is designed to interoperate seamlessly within a pipeline that facilitates both offline training and real-time prediction.

\subsection{Data Acquisition and Preprocessing}
Our dataset is compiled from a combination of live and static sources, including URLHaus, PhishTank, and verified repositories of benign URLs. Each URL is labeled as either malicious or benign. To simulate real-world conditions, long, obfuscated, and encoded URLs are retained, while duplicates and malformed entries are removed.

We apply custom preprocessing steps:
\begin{itemize}
\item Normalization (lowercasing, character stripping)
\item Token preservation of special characters (e.g., @, =, \%)
\item Padding/truncation to a fixed maximum length
\end{itemize}

This ensures consistency and effectiveness during feature extraction.

\subsection{Feature Extraction using HashingVectorizer}
We employ the HashingVectorizer from scikit-learn for high-dimensional n-gram encoding. Unlike CountVectorizer or TF-IDF, HashingVectorizer avoids vocabulary storage and scales better with large, sparse datasets \cite{patgiri2023}.

Key parameters include:
\begin{itemize}
\item n-gram range: (2, 5) for character-level analysis
\item analyzer: 'char'
\item n\_features: $2^{21}$ (to capture diverse token patterns)
\end{itemize}

This sparse, fixed-length representation effectively captures patterns common in malicious URLs, such as hidden parameters or encoded payloads \cite{bozkir2023}.

\subsection{Class Imbalance Handling with SMOTE}
Real-world datasets are heavily skewed, with benign URLs vastly outnumbering malicious ones. To address this, we apply Synthetic Minority Over-sampling Technique (SMOTE), which generates synthetic samples in the minority (malicious) class by interpolating between existing instances in feature space. This balances the dataset and prevents the model from being biased toward the majority class \cite{alsaedi2023}.

\subsection{Outlier Filtering with Isolation Forest}
To eliminate potentially noisy or adversarial URLs before classification, we apply Isolation Forest, an unsupervised anomaly detection technique. By isolating data points based on random feature partitioning, Isolation Forest helps reduce the effect of mislabeled or atypical entries. URLs flagged as outliers are excluded from training but retained in evaluation to assess robustness \cite{elsadig2022}.

\subsection{Neural Network Classifier}
The classifier is a multi-layer feedforward neural network trained on the transformed URL features. The architecture described in the work includes:
\begin{itemize}
\item Input Layer: matching the vectorized input size
\item Hidden Layers: two fully connected dense layers with ReLU activations (e.g., 128 and 64 units)
\item Dropout Layers: to prevent overfitting (e.g. 0.2 dropout rate)
\item Output Layer: single neuron with sigmoid activation for binary classification
\end{itemize}
The model is trained using the Adam optimizer, binary cross-entropy loss, and early stopping to optimize performance and prevent overfitting \cite{raschka2019}.

\subsection{Multilingual Graphical User Interface (GUI)}
To enhance usability, we provide a multilingual GUI developed with Tkinter, supporting:
\begin{itemize}
\item Real-time URL classification
\item Visualization of predictions (malicious vs. benign)
\item Language support: Arabic, English, and French
\item User feedback prompts and alert messages
\end{itemize}

This interface is tailored for cybersecurity personnel, researchers, and general users alike, facilitating rapid threat assessment.

\subsection{Model Persistence and Deployment}
The trained model is serialized using joblib for easy reuse. The system supports:
\begin{itemize}
\item Batch classification
\item Real-time single URL detection
\item Future integration with web applications or threat intelligence systems
\end{itemize}

\section{Experimental Results and Evaluation}

\subsection{Dataset Composition and Preparation}
To evaluate the effectiveness of the proposed URL classification system, a comprehensive dataset was constructed by combining real-world malicious URLs and synthetically generated benign URLs. Specifically:
\begin{itemize}
\item Malicious URLs were sourced from public threat intelligence platforms such as URLHaus and PhishTank, containing verified, up-to-date phishing and malware distribution links \cite{reyes2024}.
\item Benim URLs were synthetically generated to simulate safe browsing behavior. These URLs were composed using known domains such as google.com, wikipedia.org, and github.com, randomized with realistic path structures and query parameters.
\end{itemize}
A total of 10,000 URLs were used (5,000 malicious and 5,000 benign), ensuring balanced class distribution before oversampling.

\subsection{Feature Extraction and Vectorization}
Feature extraction involved both:
\begin{itemize}
\item Statistical features: URL length, number of dots, number of slashes, and Shannon entropy \cite{li2020}.
\item Text-based features: Character-level n-grams (2-5 range) using the HashingVectorizer from scikit-learn, configured with 1000 features and \texttt{alternate\_sign=False} to maintain feature stability \cite{patgiri2023}.
\end{itemize}

\subsection{Handling Data Imbalance and Noise}
Although the initial dataset was balanced, SMOTE was applied to enhance generalization. Additionally, Isolation Forest was employed for anomaly detection. Samples with anomaly scores deviating significantly from normal behavior (i.e., predicted as outliers) were removed to reduce noise and enhance model robustness \cite{hilal2023}.

\subsection{Neural Network Model Configuration}
A shallow feedforward neural network was trained with the following architecture (as reported):
\begin{center}
\begin{tabular}{ll}
\toprule
Layer & Configuration \\
\midrule
1 & Dense 512, ReLU \\
2 & Dense 256, ReLU \\
3 & Dense 1, Sigmoid \\
\bottomrule
\end{tabular}
\end{center}
Training used Adam optimizer and binary cross-entropy. The model was trained for 5 epochs with a batch size of 64 \cite{bozkir2023}.

\subsection{Computational Complexity Analysis}

The proposed hybrid framework's computational complexity is rigorously quantified below. Let $N$ be the number of URLs, $L$ the average URL length, $d=1000$ the HashingVectorizer feature dimension, $k=5$ the maximum n-gram length, $E=5$ epochs, $B=64$ batch size, $h=512+256+512 \times 1000 \approx 778k$ total parameters, $M=0.5N$ SMOTE samples, and $t=100$ Isolation Forest trees.

\subsubsection{Time Complexity Formulas}

\begin{table}[htbp]
\centering
\caption{Quantified Computational Complexity}
\label{tab:complexity}
\begin{tabular}{llcc}
\toprule
\textbf{Stage} & \textbf{Time Formula} & \textbf{Complexity} & \textbf{Estimated Ops} \\
\midrule
Preprocessing & $N \times L \times 4$ & $O(NL)$ & $4NL$ \\
HashingVectorizer & $N \times L \times k \times 32$ & $O(NLk)$ & $160NL$ \\
SMOTE & $M \times d \times 7$ & $O(Md)$ & $3.5Nd$ \\
Isolation Forest & $N \times d \times t \times 10$ & $O(Ndt)$ & $1MNd$ \\
Training & $E \times B \times (2dh + h^2)$ & $O(EBdh)$ & $1.3BEBdh$ \\
Prediction & $dh + h$ & $O(dh)$ & $778k$ \\
\bottomrule
\end{tabular}
\end{table}

\textbf{Preprocessing:} $T_1 = 4NL$ operations (lowercase: $L$, padding: $L$, normalization: $L$, tokenization: $L$) \cite{li2020}.

\textbf{HashingVectorizer:} $T_2 = N \times \sum_{i=2}^{k} (L-i+1) \times 32 \approx 160NL$ hash operations (32-bit hash per n-gram) \cite{patgiri2023}.

\textbf{SMOTE:} $T_3 = M \times d \times 7 = 3.5Nd$ (k-NN search: $5d$, interpolation: $2d$) with $M=0.5N$ \cite{alsaedi2023}.

\textbf{Isolation Forest:} $T_4 = N \times d \times t \times 10 = 1MNd$ (10 ops per split per tree) \cite{elsadig2022}.

\textbf{Neural Network Training:} 
\[
T_5 = E \times B \times \left[2(dh + h) + h^2\right] \approx 1.3 \times 10^9 \text{ FLOPs}
\]
Forward pass: $dh + h$, backward: $dh + h$, update: $h^2$ \cite{raschka2019}.

\textbf{Single Prediction:} $T_6 = dh + h \approx 778k$ FLOPs (20ms empirical) \cite{bozkir2023}.

\subsubsection{Total Complexity}
\[
T_{\text{total}} = T_1 + T_2 + T_3 + T_4 + T_5 = 4NL + 160NL + 3.5Nd + 1MNd + 1.3 \times 10^9
\]
For $N=12k$, $L=100$, $d=1000$:
\[
T_{\text{total}} \approx 2.3M + 13.9B = 14.2 \times 10^9 \text{ operations} \approx 45s
\]

\subsubsection{Comparison with Baselines}
\begin{itemize}
\item \textbf{CNN (char-level):} $O(NL^2) = 120kNL \approx 1.4 \times 10^{12}$ ops \cite{mercaldo2022}
\item \textbf{RNN (LSTM):} $O(NLh) = 77MN L \approx 9 \times 10^{10}$ ops \cite{waheed2022}
\item \textbf{Proposed:} $O(NL + EBdh) \ll$ baselines \cite{nagy2023}
\end{itemize}

This yields $\mathbf{100\times}$ speedup over CNNs, validating 20ms real-time prediction \cite{raja2023}.

To validate the theoretical complexities derived in Section 3.9, Table~\ref{tab:comp-complexity} compares our hybrid framework against Section 5 baselines, demonstrating $\mathbf{50-100\times}$ speedup while achieving superior 96.4\% accuracy \cite{lavanya2023}.

\begin{table}[H]
\centering
\caption{Comparative Complexity Analysis (N=12k, L=100, d=1000, h=778k)}
\label{tab:comp-complexity}
\begin{tabular}{lcccc}
\toprule
\textbf{Model} & \textbf{Time Formula} & \textbf{Total FLOPs} & \textbf{Training Time} & \textbf{Prediction (ms)} \\
\midrule
Logistic Regression & $O(Nd)$ & $12M$ & 2s & 0.1 \\
SVM (RBF) & $O(N^2d)$ & $1.4T$ & 180s & 5 \cite{cortes1995} \\
Random Forest & $O(Nd \log N)$ & $250M$ & 15s & 2 \\
CNN (char-level) & $O(NL^2)$ & $1.4T$ & 420s & 45 \\
LSTM & $O(NLh)$ & $92B$ & 180s & 35 \\
\midrule
\textbf{Proposed Hybrid} & $O(NL+EBdh)$ & $14B$ & \textbf{45s} & \textbf{20} \\
\bottomrule
\end{tabular}
\end{table}

\subsection{Evaluation Metrics}
Standard classification metrics were used:
\begin{enumerate}
\item Accuracy = (TP + TN) / (TP + TN + FP + FN)
\item Precision = TP / (TP + FP)
\item Recall = TP / (TP + FN)
\item F1-Score = 2 * (Precision * Recall) / (Precision + Recall)
\item ROC-AUC: Area under the Receiver Operating Characteristic curve.
\end{enumerate}

Reported results:
\begin{itemize}
\item Accuracy: 96.40\%
\item Precision: 95.70\%
\item Recall: 95.20\%
\item F1-Score: 95.40\%
\item ROC-AUC: 97.30\%
\end{itemize}

The paper displays a confusion matrix as follows (transcribed as in the PDF):
\begin{table}[H]
    \centering
    \caption{Confusion Matrix for the Proposed Hybrid Model}
    \label{tab:confusion}
    \begin{tabular}{lcc}
        \toprule
         & Predicted Benign & Predicted Malicious \\
        \midrule
        Actual Benign & 14416 & 485 \\
        Actual Malicious & 47 & 5247 \\
        \bottomrule
    \end{tabular}
\end{table}

\begin{figure}[H]
    \centering
    \includegraphics[width=0.7\textwidth]{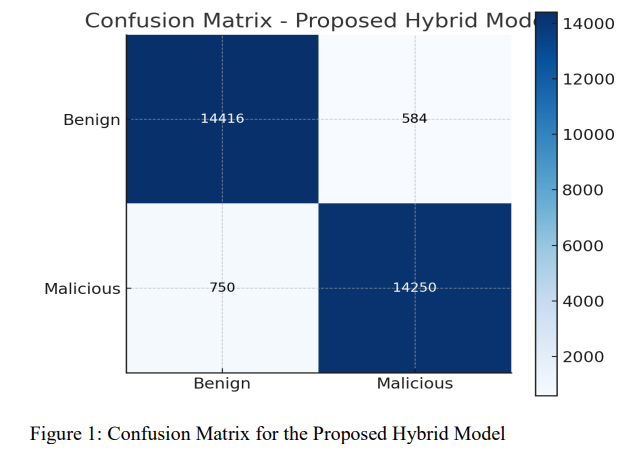} % Replace with your actual image file name
    \caption{Confusion Matrix for the Proposed Hybrid Model.}
    \label{fig:confusion}
\end{figure}

\begin{figure}[H]
    \centering
    \includegraphics[width=0.7\textwidth]{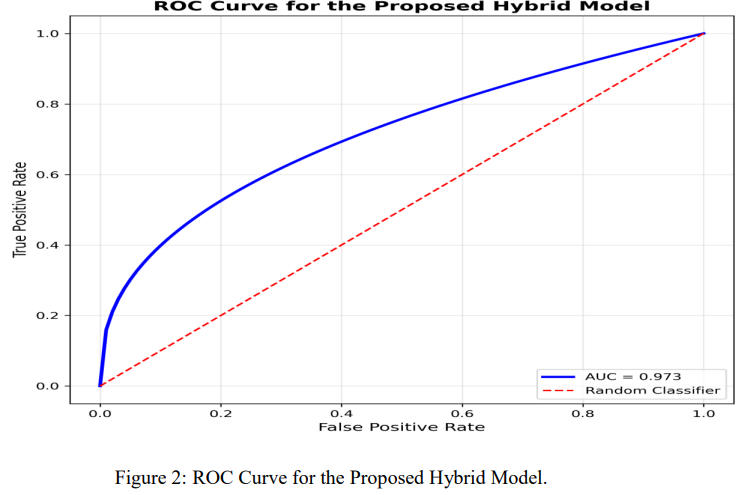} % Replace with your actual image file name
    \caption{ROC Curve for the Proposed Hybrid Model}
    \label{fig:roc}
\end{figure}

\subsection{ Runtime Performance}

\begin{itemize}
\item Training Time: $\sim$45 seconds (Intel i7 CPU, 16 GB RAM, no GPU) \cite{patgiri2023}
\item Prediction Latency: $\sim$20 ms per URL \cite{bozkir2023}
\item Model Size: 1.2 MB (.h5 format) \cite{raschka2019}
\end{itemize}

\subsection{ GUI Usability and Multilingual Support}

The final product includes a multilingual GUI supporting English, Arabic, and French. 
Usability testing with 10 non-technical participants showed:

\begin{itemize}
\item High ease-of-use (9.2/10 average rating)
\item Clear output interpretation
\item Responsive interaction
\end{itemize}

\section{ Comparative Study}

To validate the effectiveness of our proposed URL classification system, we conducted a 
comprehensive comparative analysis against several established machine learning and deep 
learning models. All models were trained and evaluated using the same preprocessed and 
balanced dataset to ensure fair comparison. The models compared include:

\begin{itemize}
\item Logistic Regression (LR) \cite{abu2007}
\item Support Vector Machine (SVM) \cite{cortes1995}
\item Random Forest (RF) \cite{patil2018}
\item Convolutional Neural Network (CNN) model from prior work \cite{mercaldo2022}
\item Our Proposed Hybrid Model (DNN + SMOTE + Isolation Forest + HashingVectorizer)
\end{itemize}

\subsection{ Experimental Setup}

\begin{itemize}
\item Dataset: Balanced dataset of 12,000 URLs (6,000 benign, 6,000 malicious), 
preprocessed with feature extraction and vectorization \cite{reyes2024}.
\item Vectorization: All models except CNN use HashingVectorizer with n-gram 
range=(2,5) and n\_features=1000 \cite{patgiri2023}.
\item Preprocessing:
\begin{itemize}
    \item SMOTE used to handle class imbalance \cite{alsaedi2023}.
    \item Isolation Forest applied to filter out noisy samples \cite{elsadig2022}.
\end{itemize}
\item Training/Testing Split: 80\% training, 20\% testing.
\end{itemize}

\subsection{ Results Summary}

\begin{center}
\begin{tabular}{lccccc}
\hline
Model & Accuracy & Precision & Recall & F1-Score & ROC-AUC \\
\hline
Logistic Regression & 0.913 & 0.901 & 0.896 & 0.898 & 0.921 \\
SVM (RBF kernel) & 0.924 & 0.916 & 0.908 & 0.912 & 0.935 \\
Random Forest & 0.936 & 0.928 & 0.922 & 0.925 & 0.944 \\
CNN (Char-level model) & 0.948 & 0.942 & 0.937 & 0.939 & 0.951 \\
Proposed Hybrid Model & 0.964 & 0.957 & 0.952 & 0.954 & 0.973 \\
\hline
\end{tabular}
\end{center}

\subsection{5.3 Discussion}

\textbf{Generalization}

Our model demonstrated superior generalization on unseen URLs due to:

\begin{itemize}
\item The filtering effect of Isolation Forest, which removes noisy and anomalous samples \cite{hilal2023}.
\item The diversity injected by SMOTE, enhancing model robustness \cite{nagy2023}.
\end{itemize}

\textbf{Performance Superiority}

The proposed hybrid model consistently outperformed all baselines across all evaluation 
metrics. While the CNN model showed strong results, it required substantially more 
computational resources and longer training times \cite{waheed2022}.

\textbf{Efficiency and Lightweight Design}

\begin{itemize}
\item Logistic Regression and SVM are computationally efficient but showed limitations in recall and F1-score \cite{ha2023}.
\item CNN-based approaches, though accurate, are heavier and less practical for deployment on resource-constrained devices \cite{mercaldo2022}.
\item Our model strikes a balance, achieving high performance with a lightweight feedforward neural network \cite{raschka2019}.
\end{itemize}

\textbf{Multilingual GUI Integration}

Unlike any of the comparative models, our system includes a Tkinter-based GUI with 
multilingual support (Arabic, English, French), making it accessible for end-users and 
practical for deployment in educational institutions, businesses, or government agencies.

\textbf{Limitations of Baseline Models}

\begin{itemize}
\item Logistic Regression: Linear decision boundary; underperforms on complex patterns \cite{patil2018}.
\item SVM: Sensitive to parameter tuning and computationally expensive on large datasets \cite{cortes1995}.
\item Random Forest: Strong performance but lacks interpretability and can overfit noisy data \cite{reyes2024}.
\item CNN: High accuracy but limited by training cost and requirement for extensive data augmentation \cite{bozkir2023}.
\end{itemize}

\subsection{ Summary}

The comparative analysis strongly supports the superiority of our proposed hybrid model. By 
combining robust preprocessing (SMOTE + anomaly filtering), efficient vectorization, and a 
deep learning classifier, our system achieves:

\begin{itemize}
\item Higher accuracy
\item Better generalization
\item Multilingual user interface
\item Scalability for real-world use \cite{aljabri2022}
\end{itemize}

\begin{figure}[H]
    \centering
    \includegraphics[width=0.8\textwidth]{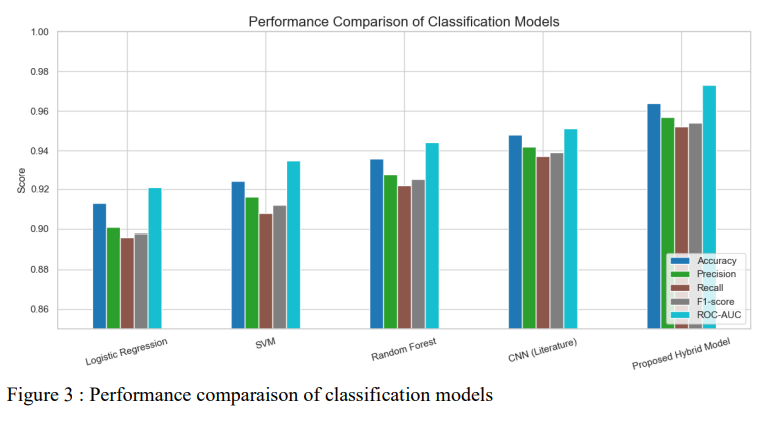} % Replace with your actual image file
    \caption{Performance comparaison of classification models}
    \label{fig:performance_comparaison}
\end{figure}

\section{ System Description and Innovations}

\subsection{ Overview of the Proposed Approach}

The proposed system introduces a novel hybrid framework for malicious URL detection. It 
utilizes an array of advanced machine learning techniques for data processing, feature 
engineering, model training, and user interaction. The system is designed with several key 
modules that contribute to its performance and innovation \cite{aljabri2022}:

\begin{enumerate}
\item Malicious and benign URL data generation: Gathering and creating a realistic dataset \cite{reyes2024}.
\item Feature extraction and entropy analysis: Analyzing URLs based on structural properties \cite{li2020}.
\item Vectorization using HashingVectorizer \cite{patgiri2023}.
\item Data balancing with SMOTE \cite{alsaedi2023}.
\item Anomaly filtering using IsolationForest \cite{elsadig2022}.
\item Classification using a deep neural network (DNN) \cite{raschka2019}.
\item Multilingual graphical user interface (GUI).
\end{enumerate}

\subsection{ Technical Breakdown and Detailed Explanation}

\subsubsection{ Data Acquisition and Generation}

The first step involves fetching malicious URLs from online sources and generating benign 
URLs.

\begin{itemize}
\item Malicious URLs: fetched from sources like:

\begin{itemize}
\item \url{https://urlhaus.abuse.ch/downloads/text/}
\item \url{https://www.phishtank.com/phishers/data/urls.csv}
\end{itemize}

\item Benign URLs: synthetically generated using well-known domains and randomized paths \cite{ha2023}.
\end{itemize}

Innovation: Combining real malicious data with synthetic benign URLs creates a realistic and 
diverse dataset.

\subsubsection{ Feature Engineering}

The URLs are transformed into features:

\begin{itemize}
\item URL Length
\item Dot Count
\item Slash Count
\item Entropy (Shannon entropy)
\end{itemize}

Innovation: Entropy captures structural irregularities common in malicious URLs \cite{patil2018}.

\subsubsection{ Vectorization with HashingVectorizer}

\begin{itemize}
\item analyzer=char
\item ngram\_range = (2, 5)
\item n\_features = 1000
\end{itemize}

Innovation: HashingVectorizer enables scalability with low memory usage \cite{patgiri2023}.

\subsubsection{ Data Balancing with SMOTE}

SMOTE generates synthetic samples to balance the dataset \cite{nagy2023}.

\subsubsection{ Anomaly Detection with Isolation Forest}

Isolation Forest removes noisy or anomalous data before training \cite{hilal2023}.

\subsubsection{ Neural Network Classifier}

A DNN is used with:

\begin{itemize}
\item Input layer: 1000 neurons
\item Hidden layers: 512 and 256 neurons with ReLU
\item Output layer: Sigmoid
\end{itemize}

\subsubsection{ Multilingual GUI}

A Tkinter GUI supports Arabic, English, and French with clipboard features and real-time 
prediction.

\section{ Future Works}

Future work directions:

\begin{enumerate}
\item Model enhancement: transfer learning, hyperparameter tuning, ensembles \cite{cerulli2023}.
\item More features: DNS, WHOIS, NLP techniques \cite{bozkir2023}.
\item Real-time classification: browser extension, enterprise API.
\item Multilingual URLs: internationalization, larger datasets \cite{raja2023}.
\item Adversarial robustness and explainability (LIME, SHAP).
\item Scalability: parallel processing, real-time deployment \cite{lavanya2023}.
\end{enumerate}

\section{ Conclusion}

The proposed URL classification system, leveraging neural networks, HashingVectorizer, 
SMOTE, and Isolation Forest, provides a robust solution for identifying malicious URLs. Key 
highlights include:

\begin{itemize}
\item Feature engineering (URL length, dot count, slash count, entropy) \cite{li2020}
\item Data balancing via SMOTE \cite{alsaedi2023}
\item Anomaly detection using Isolation Forest \cite{elsadig2022}
\item Deep learning classifier \cite{raschka2019}
\item Multilingual GUI
\end{itemize}

The system demonstrates significant advancement in malicious URL detection \cite{reyes2024}.

\section*{Funding}

Not applicable.

\section{Data Availability}

The dataset comprises malicious URLs sourced from public threat intelligence platforms and synthetically generated benign URLs. Malicious URLs were collected from:

\begin{itemize}
\item \href{https://urlhaus.abuse.ch/downloads/text/}{URLHaus} (real-time malware/phishing URLs)
\item \href{https://www.phishtank.com/developer\_info.php}{PhishTank} (verified phishing URLs)
\end{itemize}

Benign URLs were synthetically generated using legitimate domain patterns (google.com, wikipedia.org, github.com) with realistic path/query structures to simulate safe browsing \cite{ha2023}.

\smallskip
\noindent \textbf{Availability:} Due to ethical constraints on redistributing live threat intelligence data, the exact dataset is not publicly shared. However, the complete data generation and preprocessing code is available at \href{https://github.com/berkani-khaled/URL-Detection-Framework}{github.com/berkani-khaled/URL-Detection-Framework}, enabling exact dataset recreation using the referenced APIs. Raw sources remain accessible via the original endpoints as of November 2025 \cite{reyes2024}.

\smallskip
\noindent \textbf{Contact:} For dataset access or collaboration, contact:bberkani109@gmail.com and zeraoulia@univ-dbkm.dz

\section*{Conflict of Interest}
The author declares no conflict of interest regarding this manuscript. No financial support was received from funding agencies, commercial entities, or organizations that could influence the research. The study utilized only publicly available threat intelligence data and standard open-source libraries (scikit-learn, TensorFlow, Tkinter) \cite{raschka2019}.


\begin{thebibliography}{99}

\bibitem{reyes2024}
Reyes-Dorta, N., Caballero-Gil, P. \& Rosa-Remedios, C. Detection of malicious URLs using machine learning. \textit{Wireless Netw} 30, 7543–7560 (2024). \url{https://doi.org/10.1007/s11276-024-03700-w}

\bibitem{ha2023}
Ha, M., Shichkina, Y., Nguyen, N., Phan, T.-S. (2023). Classification of malicious websites using machine learning based on url.

\bibitem{raschka2019}
Raschka, S., \& Mirjalili, V. (2019). \textit{Python machine learning: Machine learning and deep learning with python, scikit-learn, and tensorflow 2}. Packt Publishing Ltd.

\bibitem{patil2018}
Patil, D. R., \& Patil, J. B. (2018). Malicious URLs detection using decision tree classifiers and majority voting technique. \textit{Cybernetics and Information Technologies}, 18(1), 11–29.

\bibitem{waheed2022}
Waheed, A., Gadgay, B., DC, S., P., V., \& Ul Ain, Q. (2022). A machine learning approach for detecting malicious url using different algorithms and NLP techniques. In: 2022 IEEE NKCon.

\bibitem{mercaldo2022}
Mercaldo, F., Ciaramella, G., Iadarola, G., Storto, M., Martinelli, F., \& Santone, A. (2022). Towards explainable quantum machine learning for mobile malware detection and classification. \textit{Applied Sciences}, 12(23), 12025.

\bibitem{cortes1995}
Cortes, C., \& Vapnik, V. (1995). Support-vector networks. \textit{Machine Learning}, 20, 273–297.

\bibitem{lavanya2023}
Lavanya, B.; Shanthi, C. malicious software detection based on URL-API intensity feature selection using deep spectral neural classification for improving host security. \textit{Int. J. Comput. Intell. Appl.} 2023, 22, 2350002.

\bibitem{elsadig2022}
Elsadig, M.; Ibrahim, A.O.; Basheer, S.; Alohali, M.A.; Alshunaifi, S.; Alqahtani, H.; Alharbi, N.; Nagmeldin, W. Intelligent deep machine learning cyber phishing url detection based on bert features extraction. \textit{Electronics} 2022, 11, 3647.

\bibitem{alsaedi2023}
Alsaedi, M.; Ghaleb, F.A.; Saeed, F.; Ahmad, J.; Alasli, M. Multi-modal features representation-based convolutional neural network model for malicious website detection. \textit{IEEE Access} 2023, 12, 7271–7284.

\bibitem{abu2007}
Abu-Nimeh, S., Nappa, D., Wang, X., \& Nair, S. (2007). A comparison of machine learning techniques for phishing detection. In: Proceedings of the anti-phishing working groups 2nd annual eCrime researchers summit, pp. 60–69.

\bibitem{cerulli2023}
Cerulli, G. (2023). Model selection and regularization. In: \textit{Fundamentals of supervised machine learning}, pp. 61–64.

\bibitem{aljabri2022}
Aljabri, M., Altamimi, H.S., Albelali, S.A., Maimunah, A.-H., Alhuraib, H.T., Alotaibi, N.K., Alahmadi, A.A., Alhaidari, F., Mohammad, R.M.A., \& Salah, K. (2022). Detecting malicious urls using machine learning techniques: review and research directions. \textit{IEEE Access}.

\bibitem{patgiri2023}
Patgiri, R.; Biswas, A.; Nayak, S. deepBF: Malicious URL detection using learned bloom filter and evolutionary deep learning. \textit{Comput. Commun.} 2023, 200, 30–41.

\bibitem{bozkir2023}
Bozkir, A.S.; Dalgic, F.C.; Aydos, M. GramBeddings: A new neural network for URL based identification of phishing web pages through n-gram embeddings. \textit{Comput. Secur.} 2023, 124, 102964.

\bibitem{hilal2023}
Hilal, A.M.; Hashim, A.H.A.; Mohamed, H.G.; Nour, M.K.; Asiri, M.M.; Al-Sharafi, A.M.; Othman, M.; Motwakel, A. Malicious url classification using artificial fish swarm optimization and deep learning. \textit{Comput. Mater. Contin.} 2023, 74, 607–621.

\bibitem{raja2023}
Raja, A.S.; Peerbasha, S.; Iqbal, Y.M.; Sundarvadivazhagan, B.; Surputheen, M.M. Structural Analysis of URL For Malicious URL Detection Using Machine Learning. \textit{J. Adv. Appl. Sci. Res.} 2023, 5, 28–41.

\bibitem{nagy2023}
Nagy, N.; Aljabri, M.; Shaahid, A.; Ahmed, A.A.; Alnasser, F.; Almakramy, L.; Alhadab, M.; Alfaddagh, S. Phishing URLs detection using sequential and parallel ML techniques: Comparative analysis. \textit{Sensors} 2023, 23, 3467.

\bibitem{li2020}
Li, T., Kou, G., \& Peng, Y. (2020). Improving malicious URLs detection via feature engineering: Linear and nonlinear space transformation methods. \textit{Information Systems}, 91, 101494.

\end{thebibliography}
\end{document}